\documentclass[aps,showpacs,preprintnumbers,amsmath,amssymb,a4paper,preprint]{revtex4}
\usepackage{graphicx}
\usepackage{dcolumn}
\usepackage{bm}
\usepackage{color}

\begin{document}

\title{Photoelectron circular dichroism in the multiphoton ionization by short laser pulses:
II. Three- and four-photon ionization of fenchone and camphor}

\author{\firstname{Anne~D.} \surname{M\"{u}ller}}
\affiliation{Institute of Physics and CINSaT, University of Kassel, Heinrich-Plett-Str.~40, 34132 Kassel, Germany}

\author{\firstname{Anton~N.} \surname{Artemyev}}
\affiliation{Institute of Physics and CINSaT, University of Kassel, Heinrich-Plett-Str.~40, 34132 Kassel, Germany}

\author{\firstname{Philipp~V.} \surname{Demekhin}}\email{demekhin@physik.uni-kassel.de}
\affiliation{Institute of Physics and CINSaT, University of Kassel, Heinrich-Plett-Str.~40, 34132 Kassel, Germany}

\date{\today}

\begin{abstract}
Angle-resolved multiphoton ionization of fenchone and camphor by short intense laser pulses is computed by the time-dependent Single Center (TDSC) method. Thereby, the photoelectron circular dichroism (PECD) in the three-photon resonance enhanced ionization and four-photon above-threshold ionization of these molecules is investigated in detail. The computational results are in a satisfactory agreement with the available experimental data, measured for randomly-oriented fenchone and camphor molecules at different wavelengths of the exciting pulses. We predict a significant enhancement of the multiphoton PECD for uniaxially-oriented fenchone and camphor.
\end{abstract}


\maketitle

\section{Introduction}
\label{sec:Intro}

The photoelectron circular dichroism (PECD) is a fascinating phenomenon of the forward-backward asymmetry in the emission of photoelectrons from randomly-oriented chiral molecules irradiated by circularly polarized light \cite{Define}. It was first predicted theoretically  \cite{Ritchie1,Ritchie2,Ritchie3} and after about a quarter of century verified in experiments with circularly polarized synchrotron radiation on bromocamphor  \cite{CDwf01} and camphor \cite{CDwf02} molecules in the gas phase. Contrary to the conventional circular dichroism in the absorption spectra of chiral molecules, the PECD emerges in the electric-dipole approximation \cite{Ritchie1,Ritchie2,Ritchie3}, which makes the effect particularly large. At present, it is confirmed by numerous experimental and theoretical studies, that in the one-photon ionization of randomly-oriented chiral molecules PECD exists on the scale of up to 20\%. Most of the studies of the one-photon ionization of chiral molecules are reviewed in Refs.~\cite{REV1,REV2,REV3}.

Recently \cite{Lux12AngChm,Lehmann13jcp}, a considerable PECD in the multiphoton ionization of chiral molecules by intense short circularly polarized laser pulses was demonstrated experimentally. In Ref.~\cite{Lux12AngChm}, a PECD of about 10\% was directly observed in the velocity map images of  photoelectrons after a 2+1 resonance-enhanced multiphoton ionization (REMPI) of camphor and fenchone in the gas phase. The  results on camphor were reproduced in the independent experiments by different authors \cite{Lehmann13jcp,Ram13EPJ} utilizing a coincidence detection technique. A full comprehensive account of the experimental data on the 2+1 REMPI of camphor and fenchone, extended also to norcamphor, is reported in Refs.~\cite{Lux15CPC,LUXdiss} . A detailed quantification of the multiphoton PECD is reported in Refs.~\cite{Lux15CPC,LUXdiss,Janssen14}.

At present, several aspects of the multiphoton PECD in chiral molecules are under investigation.  For instance, considerable PECD effects in the four-photon above-threshold ionization (ATI) spectra of bicyclic ketones are reported in Ref.~\cite{Lux16ATI}. The results on fenchone are confirmed by independent experiments in Ref.~\cite{Beaulieu16NJP}. The latter work discusses the universality of PECD in the multiphoton, ATI, and tunnel ionization regimes, uncovering also contributions from the HOMO(-1) state. Reference \cite{Kastner16ee} demonstrates a possibility of determination of the enantiomeric excess by means of multiphoton PECD with a sub-percent accuracy. The wavelength dependence of the multiphoton PECD, with the aim to access different intermediate two-photon resonant states in the REMPI process, is reported in Refs.~\cite{Rafiee16wl,Kastner17wl}. {zRecent} time-resolved studies of PECD demonstrate an important role of the femtosecond nuclear dynamics \cite{Beaulieu16td,Comby16td} and the attosecond electron dynamics \cite{Beaulieu17as} accompanying multiphoton ionization of chiral molecules. {  Finally, the very recent work \cite{Beaulieu18PXCD} illustrates that multiphoton PECD can be tailored by chiral excitation dynamics below the ionization threshold.}

Theoretical interpretation of the multiphoton PECD is a complicated task \cite{Lehmann13jcp,Dreissigacker14,Goetz17}. In Ref.~\cite{Dreissigacker14}, a qualitative insight into the phenomenon was obtained  with the help of the strong-field approximation by introducing Born corrections to the scattered electron continuum waves. Ref.~\cite{Lehmann13jcp} utilized the following two-step  model to describe REMPI of camphor. At first, the two-photon resonant excitation prealigns the initially randomly-oriented molecules by a preferred  excitation at selected orientations. This was assumed to be governed by the molecular polarizability tensor. Subsequently, one-photon ionization of the accordingly populated intermediate state was treated by the CMS-X$\alpha$ method of Ref.~\cite{POWISmeth}. The theoretical approach of  Ref.~\cite{Lehmann13jcp} yielded an overall qualitative agreement between the computed and measured multiphoton PECD of camphor. { A different perturbative two-step model was able to interpret a photoexcitation circular dichroism  in fenchone \cite{Beaulieu18PXCD}.}

The theoretical approach of different authors \cite{Goetz17} arrived at a semi-quantitative agreement with the experimental results on the multiphoton PECD in camphor and fenchone and, on the base of this agreement, classified intermediate electronic resonances and quantified their contributions to the observed effect. Methodologically, it utilized the two-step model from Ref.~\cite{Lehmann13jcp}. On the one hand, Ref.~\cite{Goetz17} considerably improved the theoretical description of the first two-photon coherent excitation step by introducing the two-photon absorption tensor. On the other hand, it employed hydrogenic waves to represent the electron continuum, which do not account for multiple scattering effects in the realistic chiral potential of a molecular ion in the second one-photon ionization step.

In our previous work \cite{TDSC} (hereafter referred to as paper~I), the time-dependent formulation of the Single Center method \cite{Demekhin11,Galitskiy15} for the theoretical description of the electron continuum spectrum in molecules \cite{Demekhin0910,Demekin10b,Knie14,Knie16,Ilchen17,Tia17} was developed and implemented in a computer code. The Time-Dependent Single Center (TDSC) method \cite{TDSC} consists in the propagation of single-active-electron wave packets in the molecular potentials, and it includes the dipole interaction of active electrons with laser fields nonperturbatively. In paper~I, TDSC method was applied to study PECD in the one-photon
ionization and two-photon ATI of a model methane-like chiral system by  short intense circularly polarized high-frequency laser pulses.

In the present work we apply this method to study the multiphoton ionization of the prototypical enantiomers of camphor and fenchone, for which one-photon PECD \cite{CDwf01,CDwf02,Janssen14,Hergenhahn04,Nahon06,Powis08,Nahon10,HHGPECD,Nahon16} and multiphoton PECD \cite{Lux12AngChm,Lehmann13jcp,Ram13EPJ,Lux15CPC,LUXdiss,Lux16ATI,Beaulieu16NJP,Kastner16ee,Kastner17wl,Beaulieu16td,Comby16td,Beaulieu17as,Dreissigacker14,Goetz17} have  been studied extensively. In particular, we aim at the theoretical interpretation of the experimental PECD in the three-photon REMPI \cite{Lux12AngChm,Lux15CPC,LUXdiss,Kastner17wl} and four-photon ATI \cite{LUXdiss,Lux16ATI} processes of these bicyclic ketones. The paper is organized as follows. The present computational details are outlined in Sec.~\ref{sec:Theory}. The computed  multiphoton PECD spectra of fenchone and camphor are discussed in Sec.~\ref{sec:results}. We conclude in Sec.~\ref{sec:Summary} with a brief summary and outlook.

\section{Computational details}
\label{sec:Theory}

The present theoretical approach is described in detail in paper~I. Therefore, only essential points of the TDSC method, relevant to the present study, are summarized  below. Briefly, in order to solve the time-dependent Schr\"{o}dinger equation for an active electron interacting with a molecular ion and an external laser pulse, we employ the efficient numerical approach of Refs.~\cite{Demekhin13H,Artemyev16He1,Artemyev17He2,Artemyev167He3}. The one-particle wave function of the active electron is represented in the TDSC method via the limited expansion over spherical harmonics with respect to a single molecular center. The expansion functions play the role of time-dependent wave packets with given angular momentum quantum numbers $\ell m$ (i.e., of the partial photoelectron waves). The radial coordinate is described by  a finite-element discrete-variable representation, which are covered by normalized Lagrange polynomials constructed over a Gauss-Lobatto grid \cite{appr1,appr2,appr3}. The explicit analytic expressions for the matrix elements of the Hamiltonian in this basis can be found in Refs.~\cite{Demekhin13H,Artemyev16He1,Artemyev17He2,Artemyev167He3}.

In the present work, the interaction of an active electron with laser pulses is described in the dipole-velocity gauge \cite{Artemyev17He2}, which enforces the convergence of the numerical solution over the partial waves \cite{VG1,VG2}. Calculations were performed for sine-squared laser pulses with the time-envelope $g(t)=\sin^2(\pi\frac{t}{T})$ and selected wavelengths in the range of 360--410~nm.  For computational reasons, the full propagation duration $T$ was set to 20~fs. Therefore, the present pulse duration of about 10~fs FWHM was somewhat shorter than the pulse duration of about 25~fs used in the experiments \cite{Lux12AngChm,Lux15CPC,LUXdiss,Kastner17wl,Lux16ATI}. Nevertheless, the presently considered laser pulses support about 15 optical cycles, such that any asymmetry due to the carrier-envelope phase can be neglected. The peak intensity of the pulse was set to $10^{12}$~W/cm$^2$.

In the calculations, the positions of all nuclei were fixed at the equilibrium internuclear geometry of the ground electronic state of the neutral molecules  Ref.~\cite{Goetz17}. Apart from the nuclei-electron interaction, the present potential incorporates the electrostatic Coulomb interaction of the photoelectron with all electrons in the occupied shells of the molecular ion. The latter orbitals were generated by the PC GAMESS (General Atomic and Molecular Electronic Structure System) version Alex A. Granovsky
(www http://classic.chem.msu.su/gran/gamess/index.html) of the GAMESS (US) QC package \cite{Schmidt93} within the Hartree-Fock approximation in a triplet zeta valence (TZV) basis set \cite{Dunning71} with 2 additional polarization functions of p-type and one function of d-type.

The Coulomb interaction of the photoelectron with the 83 electrons remaining in the HOMO-ionized fenchone/camphor consists of the local direct and nonlocal exchange potentials. In order to be able to carry out calculations with the TDSC method at reasonable computational costs, we replaced the exact nonlocal exchange interaction of the photoelectron with the remaining electrons by a local potential and implied the $X\alpha$-approximation \cite{XALP} (see also below). To additionally enforce the convergence of the SC solutions over the partial harmonics $\ell m$, the center of the molecule was set to the center of the total charge distribution of the ion, which includes contributions from all nuclei and electrons (apart from one HOMO electron which is ionized). With this choice of the molecular center, the permanent dipole moment of the molecular ion vanishes. Finally, $z^\prime$-axis of the molecular frame was set to coincide with the C=O bond of fenchone/camphor.

The single center expansion of photoelectron wave packets was restricted by partial waves with $\ell,\vert m\vert \leq 25$. The wave packets were propagated on the radial grid of $r \leq 425$~a.u., which supports electrons released by the three-photon ionization and four-photon ATI processes during the whole propagation time $T$. In the inner part of the fenchone/camphor molecule, the radial grid was divided in finite elements of different sizes, which were defined by the positions of the carbon and oxygen atoms. In the outer part, it was covered by equidistant finite elements of size 2.5~a.u. Each finite element was covered by 10 Gauss-Lobatto points. A mask function \cite{Artemyev17He2,Artemyev167He3} at the end of the grid was introduced to avoid reflections of very fast electrons from the grid boundary.

The initial electronic state in the absence of the pulse was obtained by the diagonalization of the stationary Hamiltonian matrix in the reduced space interval of $r\leq 25$~a.u. Thereby, the orbital no.~42, obtained in the present one-particle electron potential, represents the HOMO of camphor/fenchone. The absolute value of the computed one-electron binding energy of this orbital was equated with the experimental ionization potential (IP) of the respective HOMO (8.6~eV \cite{Powis08} for fenchone  and  8.7~eV \cite{IPCAMP} for camphor). For this purpose, the parameter $\alpha$ of the local exchange potential  $X\alpha$ was set to be equal to 0.85 and 0.87, respectively. For both molecules, thus computed orbital no.~42 exhibits an overlap of about $93$\% with the respective HOMO obtained in the Hartree-Fock approximation.

The propagation was performed at different orientations of the molecular frame (MF, $z^\prime$-axis defined by the C=O bond) with respect to the laboratory frame (LF, $z$-axis defined by the direction of the propagation of the laser pulses), which are given by two rotational Euler angles $\alpha$ and $\beta$. Owing to the axial symmetry of a circularly polarized pulse, integration over the third Euler angle $\gamma$, which describes the rotation of the MF around the laboratory  $z$-axis, can be performed analytically. The orientation intervals $\alpha\in[0,2\pi)$ and $\beta\in[0,\pi]$ were covered with the steps of $\Delta \alpha= \Delta \beta=0.1\,\pi$. To enable the propagation at an arbitrary orientation, the initial electronic state and the molecular potential were transformed from MF to LF.

The time-dependent electron wave packets were propagated using the short-iterative Lanczos method \cite{ALG1} with the following time-evolution operator $\hat{U}(t,t+\Delta t)=\exp\{-iP{\hat{H}(t)}P\Delta t\}$. Here, $\hat{H}(t)$ is the total Hamiltonian operator of a molecule exposed to a laser pulse, and   $P=1-\sum_\alpha \vert \phi_\alpha\rangle \langle \phi_\alpha\vert$ is a one-particle projector. The latter ensures orthogonality of the photoelectron wave packet $\Psi(\vec{r},t)$ to the set of orbitals $\left\{ \phi_\alpha\right\}$ at any time, since these orbitals cannot be populated by the evolution operator $\hat{U}(t,t+\Delta t)$. In the first line, this projector excludes  the orbitals nos.~1~--~41, which are occupied in the molecular ion, from the propagation. In order to increase an accuracy of the integration over time, all one-particle orbitals with the energy larger than 1000~a.u. were also excluded from the propagation through this projector. { Those orbitals $\left\{ \phi_\alpha\right\}$ were obtained by the diagonalization of the stationary Hamiltonian in the reduced space interval (see above).}

Finally, at the end of the pulse, the momentum distribution of the emitted photoelectrons was obtained by a Fourier transformation of the spatial wave packet. Because of the averaging  over the orientation angle $\gamma$ (see above), the resulting momentum distribution, computed for a particular molecular orientation ($\alpha,\beta$), becomes independent of the azimuthal angle $\varphi$. For a given photoelectron kinetic energy (momentum $k=\vert\vec{k}\vert$), this distribution depends only on the electron emission angle $\theta$ measured with respect to the direction of the propagation of the pulse. { It thus can be expanded over Legendre polynomials $P_L( \theta)$.} For randomly-oriented molecules (i.e., after averaging over the remaining  Euler angles $\alpha$ and $\beta$), the computed momentum distribution { is given by the well-known formulae} \cite{AD1,AD2,AD3,AD4,AD5}:
\begin{equation}
\label{eq:ADmpiTOT}
\widetilde{W}(\vec{k})=\sum_{L}  \widetilde{b}_L(k)\, P_L( \theta).
\end{equation}
Here and below, tilde is referring to the case of randomly-oriented molecules.

\section{Results and discussion}
\label{sec:results}

\subsection{R(--) fenchone at 400~nm}
\label{sec:resultsFEN400nm}

\begin{figure}
\includegraphics[scale=1.0]{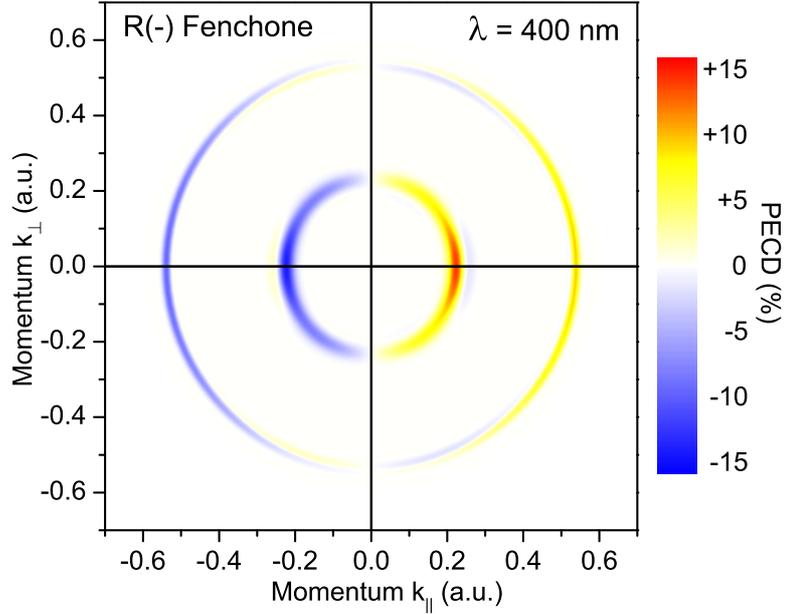}
\caption{Multiphoton PECD, computed in the present work for randomly-oriented R(--) fenchone molecules exposed to 400~nm laser pulses. The inner ring represents photoelectrons released by the three-photon ionization, while the outer ring is due to the four-photon ATI process (see also text for details). Shown is the relative difference between the two electron spectra computed for the right-handed  and for the left-handed  circular polarizations.  Each PECD signal is shown in percent relatively to the maximal intensity in the respective { PI or ATI spectrum computed for one polarization}. The pulse propagates horizontally from left to the right, i.e., the laboratory $z$-axis coincides with the $k_{||}$-axis. The computed forward-backward asymmetry in the photoemission from R(--) fenchone can directly be compared with the respective experimental PECD shown in Fig.~4.49 of Ref.~\cite{LUXdiss}.} \label{fig:fen400nm}
\end{figure}

We start the present discussion with the multiphoton ionization of randomly-oriented R(--) fenchone molecules by 400~nm laser pulses.  Fig.~\ref{fig:fen400nm} depicts the respective theoretical PECD  in the momentum space ($k_{||}$,$k_\bot$). Here, the emission angle $\theta$ is measured with respect to $k_{||}$-axis, and the forward (backward) emission directions correspond to positive (negative) values of $k_{||}$. Shown in this figure is the difference between the normalized (see below) emission distributions Eq.~({\ref{eq:ADmpiTOT}), computed for the right-handed (CPR) and left-handed (CPL) circularly polarized  pulses: $\widetilde{W}^{CPR}(\vec{k})-\widetilde{W}^{CPL}(\vec{k})$.

Two different photoelectron signals, each centered at different values of the momentum $\vert\vec{k}\vert$, are clearly seen in Fig.~\ref{fig:fen400nm}. The inner ring at  $k \approx 0.23$~a.u. represents photoelectrons (PI) released by the absorption of three photons of energy $\omega=3.10$~eV, such that $\varepsilon=k^2/2=3\omega-IP=0.70$~eV. The outer ring at $k \approx 0.53$~a.u. represents ATI photoelectrons emitted due to the four-photon absorption: $\varepsilon=4\omega-IP=3.80$~eV. The presently computed ATI spectrum is about 10 times weaker than the PI spectrum. Therefore, for a better comparison, the PECDs for the PI and ATI electrons are shown in this figure in percent of the maximal intensity,
{ respectively, in the PI and ATI spectra computed for one of the circular polarizations.}

{\linespread{1.1}
\begin{table}
\caption{The presently computed and measured odd $\widetilde{b}^{CPR}_{2n+1}$ Legendre expansion coefficients  (relative to $\widetilde{b}^{CPR}_{0}$) for randomly-oriented camphor and fenchone molecules together with the respective total $\widetilde{PECD}$  of  Eq.~(\ref{eq:mpiPECDtot}).}
\begin{ruledtabular}
\begin{tabular}{lrrrrr}
Source/Case &$\widetilde{b}^{CPR}_1$&$\widetilde{b}^{CPR}_3$&$\widetilde{b}^{CPR}_5$&$\widetilde{b}^{CPR}_7$&$\widetilde{PECD}$\\
\hline
\\ \multicolumn{3}{l}{R(--) Fenchone at 400~nm }\\ \cline{1-3}
Theor. PI&  +0.049&--0.006&+0.006&+0.002& +10.2\%  \\
Expt.$^a$ \,PI &  +0.048&--0.005&--0.003&+0.001& +9.7\% \\
{Expt.$^b$ \,PI }& {+0.055}&{--0.010}&{+0.004}& &{+11.6\%} \\
Theor. ATI& +0.026&+0.019&--0.010&+0.002& +4.0\%\\
Expt.$^a$  \,ATI &  +0.016&--0.018&+0.003&+0.005& +4.1\%\\
{Expt.$^b$  \,ATI }&{+0.027}&{--0.033}&{+0.006}& &{+7.2\%}\\
\\ \multicolumn{3}{l}{R(--) Fenchone at 380~nm }\\ \cline{1-3}
Theor. PI& --0.040&+0.021&--0.023&+0.004& --9.7\%  \\
Expt.$^c$ \,PI &  --0.014&--0.004& & & --2.6\% \\
Theor. ATI&+0.043&--0.040&+0.001&--0.001& +10.6\%\\
\\ \multicolumn{3}{l}{R(+) Camphor at 360~nm}\\ \cline{1-3}
Theor. PI& +0.022 &--0.026&--0.003&--0.005&+5.7\%\\
Expt.$^d$  \,PI &+0.019&--0.039&--0.001&+0.004& +5.7\%\\
Theor. ATI& +0.009&--0.092&+0.034&--0.010&+7.4\%\\
Expt.$^d$  \,ATI&+0.018&--0.049&+0.029&--0.004& +6.8\%\\
\end{tabular} \label{tab:propt}
\end{ruledtabular}
\par $^a$ {\footnotesize Data from Ref.~\cite{Lux16ATI} acquired at 398~nm (Table SI2 in the Supporting Information);}
\par {$^b$ {\footnotesize Data from Ref.~\cite{Beaulieu16td} acquired at 398~nm (extracted form Fig.~5 at the maxima of respective photoelectron spectra and converted from S(+) to R(--) fenchone);}}
\par $^c$ {\footnotesize Data from Ref.~\cite{Kastner17wl} (extracted from Fig.~7 for the $3p_2$ state  at the photoelectron energy of 1.21~eV and converted from S(+) to R(--) fenchone);}
\par $^d$ {\footnotesize Data from Ref.~\cite{Lux16ATI} acquired at 398~nm (Table SI1 in the Supporting Information).}
\end{table}
}

As one can see from Fig.~\ref{fig:fen400nm}, both, PI and ATI electrons released from  R(--) fenchone by the 400~nm CPR pulses, are preferably emitted in the forward direction with respect to the pulse propagation (i.e., $PECD > 0$ for $k_{||}>0$). For the three-photon ionization, the PECD reaches about 13\% of the maximal PI signal, while for the four-photon ionization it is around 9\% of the maximal ATI signal. These theoretical results are in a good agreement with the experimental results from Refs.~\cite{Lux12AngChm,Lux15CPC,LUXdiss,Lux16ATI}. The theoretical PECD from Fig.~\ref{fig:fen400nm} can directly be compared with the experimental results shown in the lower-right panel of Fig.~4.49 on page 175 in Ref.~\cite{LUXdiss}. It depicts antisymmetric component of the experimental Abel-inverted multiphoton PECD signal for R(--) fenchone exposed to intense 398~nm laser pulses. Similarly, the inner ring at $\varepsilon \approx 0.56$~eV  corresponds to the 2+1 REMPI signal, whereas outermost ring at $\varepsilon \approx 3.8$~eV to the ATI signal. { A very similar PECD signal, measured for fenchone at 398~nm, is also reported in Fig.~5 of Ref.~\cite{Beaulieu16td}.}

A further quantitative comparison of the present theoretical results, obtained for randomly-oriented R(--) fenchone molecules exposed to 400~nm pulses,  with the respective experimental results {from Refs.~\cite{Lux16ATI,Beaulieu16td}} is given in the upper part of Tab.~\ref{tab:propt}. It collects relative odd coefficient $\widetilde{b}^{CPR}_{2n+1}$ from the expansion (\ref{eq:ADmpiTOT}), computed and measured for the CPR pulse. The last column of this table summarizes respective total PECD given by a single value as \cite{Lehmann13jcp,Lux15CPC,Janssen14,Dreissigacker14}:
\begin{equation}
\label{eq:mpiPECDtot}
\widetilde{PECD}=2\widetilde{b}^{CPR}_1-\frac{1}{2}\widetilde{b}^{CPR}_3+\frac{1}{4}\widetilde{b}^{CPR}_5-\frac{5}{32}\widetilde{b}^{CPR}_7 + ...\, .
\end{equation}
Table~\ref{tab:propt} illustrates the very good agreement between the odd expansion coefficients and the total PECD, computed and measured for PI electrons. For ATI electrons, the agreement between the computed and measured expansion coefficients is satisfactory. { Unlike the present calculations, the absolute value of the measured coefficient $\widetilde{b}^{CPR}_{3}$ is larger than that of $\widetilde{b}^{CPR}_{1}$, and $\widetilde{b}^{CPR}_{3}$ is negative, resulting in an angularly structured PECD signal \cite{Lux16ATI,Beaulieu16td}.} Interestingly, because of the different signs in Eq.~(\ref{eq:mpiPECDtot}), the disagreements between the computed and measured  $\widetilde{b}^{CPR}_{1}$ coefficients and separately between the respective $\widetilde{b}^{CPR}_{3}$ coefficients compensate each other in the total PECD for ATI electrons.

\subsection{R(--) fenchone at 380~nm}
\label{sec:results380nm}

\begin{figure}
\includegraphics[scale=1.0]{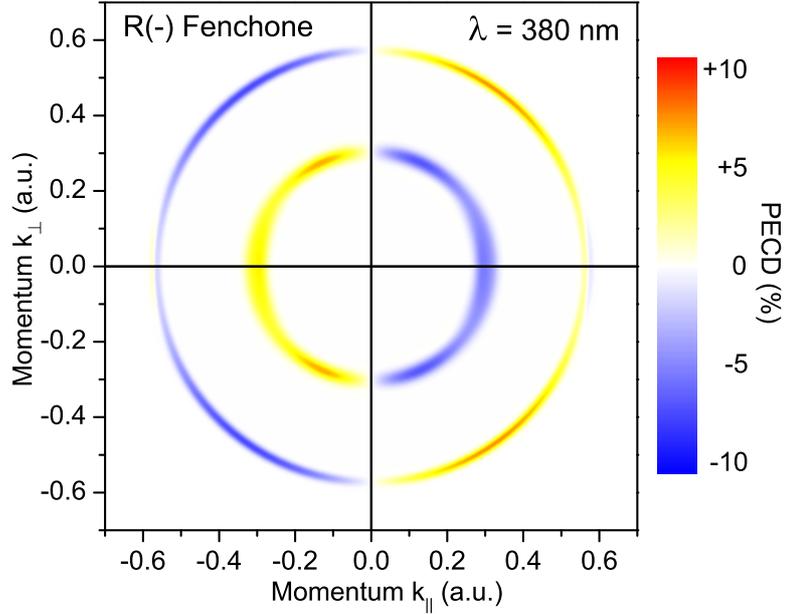}
\caption{The same as in Fig.~\ref{fig:fen400nm}, but for 380~nm laser pulses. Note that the three-photon PECD (inner ring) changed its sign compared to that obtained for 400~nm pulses (see Fig.~\ref{fig:fen400nm}).} \label{fig:fen380nm}
\end{figure}

Being confident that the present molecular potential for fenchone allows reproducing experimental results for 400~nm laser pulses, we studied the wavelength dependence of the multiphoton PECD, which was observed and characterized in Ref.~\cite{Kastner17wl}. Those calculations were performed {for longer and shorter} wavelengths in the interval of 360--410~nm {in steps of $\Delta \lambda=10$~nm}. Similar to the experimental observations reported in Ref.~\cite{Kastner17wl}, we found that the multiphoton PECD computed for the three-photon ionization of fenchone changes its sign for shorter wavelengths at about 388~nm. This fact is illustrated in Fig.~\ref{fig:fen380nm}, where the PECD computed for 380~nm laser pulses is shown in  the same way as in Fig.~\ref{fig:fen400nm} (see description in the preceding section). Here, the photoelectrons released by the absorption of three photons of energy $\omega=3.27$~eV form the inner ring at  $k \approx 0.30$~a.u. ($\varepsilon=3\omega-IP=1.21$~eV), and the outer ring at $k \approx 0.57$~a.u. is formed by four-photon ATI electrons ($\varepsilon=4\omega-IP=4.48$~eV).

By comparing the PI signals in Figs.~\ref{fig:fen400nm} and \ref{fig:fen380nm}, one can see that the computed three-photon PECD of fenchone changed its sign. A very similar situation is illustrated in Fig.~6 of Ref.~\cite{Kastner17wl}. The latter figure compares the antisymmetric components of the experimental PECD for the 2+1 REMPI of fenchone recorded at different wavelengths. The PECD acquired in Ref.~\cite{Kastner17wl} at 412~nm is very similar to that computed here for 400~nm  (Fig.~\ref{fig:fen400nm}). This signal is associated in Ref.~\cite{Kastner17wl} with the two-photon excitation of lower vibrational levels of the B-band of fenchone by the $3s \leftarrow n$ transition. Note the opposite sign of the measured PECD owing to the use of the S(+) enantiomer of fenchone in the experiments of Ref.~\cite{Kastner17wl}.

For the shorter wavelength of 376~nm, Fig.~6 of Ref.~\cite{Kastner17wl} illustrates two signals related with the three-photon ionization of fenchone. The inner ring arises from the excitation of higher vibrational levels of the intermediate B-band, while the outer ring is due to the two-photon excitation of lower vibrational levels of the intermediate C-band of fenchone via the $3p \leftarrow n$ transition. As one can see from Fig.~6 of Ref.~\cite{Kastner17wl}, the PECD measured for the excitation of the C-band at 376~nm has an opposite sign compared to that observed for B-band (similarly to the PI signal in Fig.~\ref{fig:fen380nm} computed here at 380~nm). Note, due to a frozen internuclear geometry of fenchone (see Sec.~\ref{sec:Theory}), only one (atomic-like) electronic state can be addressed in the two-photon excitation step in the present calculations.

Interestingly, the ATI signals computed in the present work for 400 and 380~nm exhibit the same signs of the four-photon PECD (cf., outer rings in Figs.~\ref{fig:fen400nm} and \ref{fig:fen380nm}). {The fact that the three- and four-photon PECDs computed for 380~nm exhibit different signs (see Fig.~\ref{fig:fen380nm}) suggests that both PI and ATI electrons are subjects to distinct multiple scattering effects.} A quantification of the present theoretical results obtained for 380~nm pulses is given in the middle of Tab.~\ref{tab:propt}. The only available experimental coefficients $\widetilde{b}^{CPR}_{1}$  and  $\widetilde{b}^{CPR}_{3}$ and the respective total $\widetilde{PECD}$, reported in Ref.~\cite{Kastner17wl} for the $3p_2$ state of the C-band, are also shown in Tab.~\ref{tab:propt} for comparison. These data were recorded for the photoelectron kinetic energy of 1.21~eV, which coincides with that of the PI signal computed at 380~nm (see above). As one can see from this table, the present theory reproduces the sign of the respective three-photon PECD but overestimates the effect  by about three times.

\begin{figure}
\includegraphics[scale=1.00]{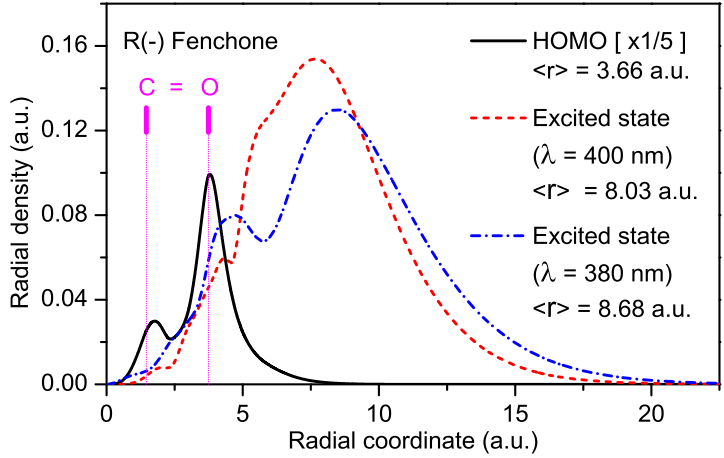}
\caption{Radial densities of the two-photon absorption intermediate electronic states of R(--) fenchone exposed to 400~nm (dashed curve) and 380~nm (dash-dotted curve) laser pulses. Radial density of the HOMO electron (solid curve) is shown  on the suppressed scale for comparison (note the $\times 1/5$ factor in the legend). Positions of the carbon and oxygen atoms from the C=O bond are also indicated by the vertical bars. The mean radii of these molecular orbitals (indicated in the legend) suggest that both excited states have a Rydberg-type character with the principal quantum number $n=3$.} \label{fig:funct}
\end{figure}

The above discussed changes of PECD for the 2+1 REMPI of fenchone with respect to the photon energy (wavelength) is associated in the literature with the effect of the two-photon excited state \cite{Lehmann13jcp,Rafiee16wl,Kastner17wl,Goetz17}. Here, the two-photon absorption matrix element introduces an orientation-dependent probability for the population of the intermediate state \cite{Goetz17} and favors in this step particular orientations of the initially randomly-oriented molecules by their preferred excitations. Indeed, the present calculations demonstrate considerable enhancements of the total photoelectron yield computed at selected molecular orientations (not shown here for brevity), which can be considered as an indirect indication of the influence of this intermediate electronic state.

In order to uncover those intermediate states, we diagonalized the stationary Hamiltonian matrix of fenchone in a larger spatial interval of $r\leq 50$ and analyzed its energy spectrum. It appears, that the absorption of two 400~nm photons excites the system from its ground electronic state to a very close proximity of the electronic eigenstate with the one-electron energy of $\varepsilon=-2.38$~eV (to be compared with $2\omega-IP=-2.40$~eV). In the case of the absorption of two 380~nm photons, the closest electronic eigenstate has the energy of $\varepsilon=-2.03$~eV (to be compared with $2\omega-IP=-2.06$~eV).   The radial densities $\vert r \Psi(r)\vert^2$ of the respective intermediate electronic states of fenchone are compared in Fig.~\ref{fig:funct} with the radial density of HOMO electron (see legend). The  mean radii of those excited states (8.03 and 8.68~a.u. for  400 and 380~nm pulses, respectively) indicate that both excited states belong to the lowest Rydberg orbitals of $3s/3p$-type.

\subsection{R(+) camphor at 360~nm}
\label{sec:resultsCAMP360nm}

Similarly to fenchone, calculations of the multiphoton ionization of camphor by short intense laser pulses were performed in the wavelength range of 360--410~nm. The three- and four-photon PECDs of camphor computed in the present work for 400~nm laser pulses (not shown here for brevity) do not reproduce experimental results of Refs.~\cite{Lux12AngChm,Lux15CPC,LUXdiss,Lux16ATI}. However, these experiments could be reproduced at a somewhat shorter wavelength. This fact is illustrated in Fig.~\ref{fig:cam360nm}, which depicts the multiphoton PECD of camphor computed for 360~nm laser pulses. Again, the inner ring at  $k\approx 0.35$~a.u. represents photoelectrons emitted by the absorption of three photons of the energy $\omega=3.45$~eV ($\varepsilon=3\omega-IP=1.65$~eV), while the outer ring at $k\approx 0.61$~a.u. is formed by the four-photon ATI ($\varepsilon=4\omega-IP=5.10$~eV).

\begin{figure}
\includegraphics[scale=1.00]{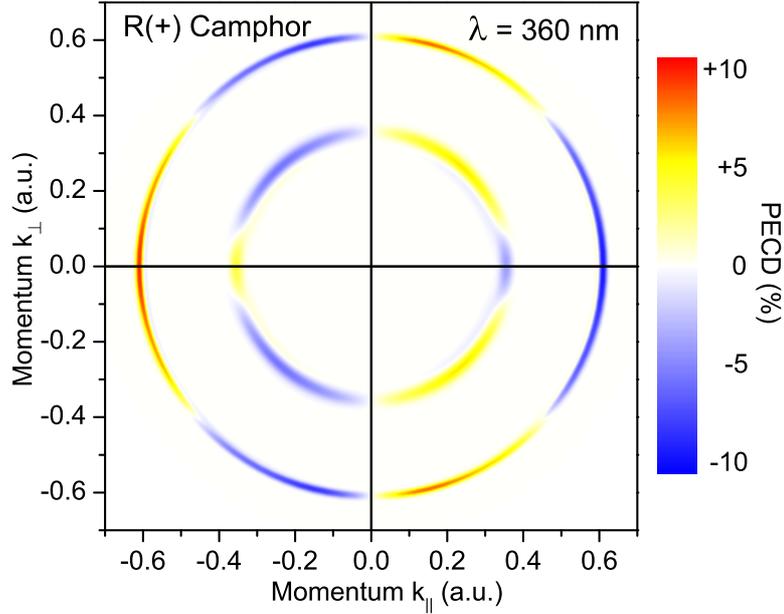}
\caption{Multiphoton PECD, computed in the present work for randomly-oriented R(+) camphor  molecules exposed to 360~nm laser pulses (see notations in Fig.~\ref{fig:fen400nm} and text for details). The presently computed forward-backward asymmetry in the photoemission from R(+) camphor can be compared with the experimental PECD spectra from Ref.~\cite{LUXdiss} (see Fig.~4.49 there). The calculations clearly reproduce nodal structures within the forward-backward hemispheres including the overall sign and size of the observed effect.} \label{fig:cam360nm}
\end{figure}

As one can see from Fig.~\ref{fig:cam360nm}, both the three- and the four-photon PECDs, computed for randomly-oriented camphor molecules exposed to 360~nm laser pulses, exhibit very similar nodal structures. In the forward hemisphere, for small emission angles the computed PECDs are negative, while for larger angles positive. The size of the computed effect is on the order of  5\% and 9\%  of the maximal PI and ATI signals, respectively. These theoretical results agree well with the experimental results from Refs.~\cite{Lux12AngChm,Lux15CPC,LUXdiss,Lux16ATI}. The computed chiral asymmetry from Fig.~\ref{fig:cam360nm} can directly be compared with the measured PECD shown in the upper-left panel of Fig.~4.49 in Ref.~\cite{LUXdiss}. Similar to the present calculations, the chiral asymmetries measured in this reference for R(+) camphor exhibit nodal structures for both PI and ATI electrons observed at the kinetic energies of $\varepsilon \approx 0.52$~eV and $3.7$~eV, respectively.

The odd Legendre expansion coefficients (\ref{eq:ADmpiTOT})  and the respective total $\widetilde{PECD}s$ (\ref{eq:mpiPECDtot}), computed in the present work for PI and ATI electrons released from randomly-oriented camphor molecules by 360~nm CPR pulses, are collected in the lower part of Tab.~\ref{tab:propt} together with the corresponding experimental quantities from Ref.~\cite{Lux16ATI}. This table demonstrates a good agreement between the data computed here for the wavelength of 360~nm and those measured in Ref.~\cite{Lux16ATI} at 398~nm. Similarly to the experimental observations, the $\widetilde{b}^{CPR}_{3}$ coefficients computed here for the PI and ATI signals provide dominant contributions to the respective total asymmetries (\ref{eq:mpiPECDtot}) and create thereby the observed nodal structures. {As in the case of fenchone, the absorption of two 360~nm photons excites camphor to a close proximity of the electronic eigenstate with the energy of $\varepsilon=-1.78$~eV (to be compared with $2 \omega-IP=-1.80$~eV, not shown in Fig.~\ref{fig:funct} for brevity).}

The presently used 360~nm laser pulses populate the electron continuum spectrum of camphor at somewhat higher kinetic energies than 400~nm pulses applied in the experiments \cite{Lux12AngChm,Lux15CPC,LUXdiss,Lux16ATI} (cf., $\varepsilon = 1.65$ and 5.10~eV for PI and ATI electrons released by 360~nm pulses versus $\varepsilon = 0.60$ and 3.7~eV  for 400~nm pulses, respectively). Good agreement between the respective theoretical and experimental  results indicates that the electron continuum spectrum of camphor, generated by the presently used molecular potential close above its ionization threshold, is slightly shifted towards larger kinetic energies. This potential neglects electron correlations and utilizes local exchange interaction, which makes it somewhat weaker. It seems that the above mentioned shortcoming of the molecular potential of fenchone compensate each other, providing an adequate attraction strength for the photoelectron in the continuum. A detailed clarification of this issue requires extended theoretical investigations which are outside the scope of the present work.

\subsection{Uniaxially-oriented molecules}
\label{sec:resultsORIENT}

As it was demonstrated in paper~I, the PECD computed for individual orientations of a model  methane-like chiral system can be larger than 50\%, and averaging over all orientations results in a considerably smaller effect. Recently, this effect was confirmed experimentally by observing a significantly enhanced chiral asymmetry in the inner-shell photoionization of uniaxially-oriented methyloxirane \cite{Tia17}. In this section, we study the effect of the molecular orientation on the multiphoton PECD of fenchone and camphor. Here, the C=O bond of both molecules naturally defines  $z^\prime$-axis, but no further molecular axes can be introduced uniquely. Therefore, in order to investigate chiral asymmetry of uniaxially-oriented fenchone and camphor, the presently computed angle-resolved multiphoton ionization spectra were  averaged over the orientation angle $\alpha$.

\begin{widetext}
Similarly to paper~I, we introduce  multiphoton PECD of the uniaxially-oriented molecules as
\begin{multline}
\label{eq:mpiPECD}
PECD(\beta)=\left(b^{CPR}_1(\beta)-\frac{b^{CPR}_3(\beta)}{4}+\frac{b^{CPR}_5(\beta)}{8}-\frac{5b^{CPR}_7(\beta)}{64} \right)  \\ - \left(b^{CPL}_1(\beta)-\frac{b^{CPL}_3(\beta)}{4}+\frac{b^{CPL}_5(\beta)}{8}-\frac{5b^{CPL}_7(\beta)}{64} \right).
\end{multline}
Because of the known asymmetry property of odd Legendre coefficients, $b^{CPL}_{2n+1}(\beta)=-b^{CPR}_{2n+1}(\pi-\beta)$ \cite{Dreissigacker14}, equation~(\ref{eq:mpiPECD}) can be reduced to the totaly-symmetric (with respect to the angle $\beta=\frac{\pi}{2}$) form
\begin{multline}
\label{eq:mpiPECDbet}
PECD(\beta)=
\left\{b^{CPR}_1(\beta)+ b^{CPR}_1(\pi-\beta) \right\} -\frac{1}{4}\left\{b^{CPR}_3(\beta)+b^{CPR}_3(\pi-\beta)\right\}\\ + \frac{1}{8}\left\{b^{CPR}_5(\beta)+b^{CPR}_5(\pi-\beta)\right\} - \frac{5}{64}\left\{b^{CPR}_7(\beta)+b^{CPR}_7(\pi-\beta)\right\}  \, .
\end{multline}
After averaging over orientation angle $\beta$, equation~(\ref{eq:mpiPECDbet}) reduces to its well-known \cite{Lehmann13jcp,Lux15CPC,Janssen14,Dreissigacker14} form of Eq.~(\ref{eq:mpiPECDtot}). Figure~\ref{fig:orient} depicts the multiphoton PECDs computed via Eq.~(\ref{eq:mpiPECDbet}) for uniaxially-oriented fenchone and camphor molecules exposed to 400~nm and 360~nm laser pulses, respectively.
\end{widetext}

For the three-photon PI signal of fenchone (solid squares in Fig.~\ref{fig:orient}), the orientation of the C=O bond along (parallel and antiparallel to) the direction of the propagation of the exciting pulses results in the effect of +19.1\% (to be compared with the total $\widetilde{PECD}$ of +10.2\%). For the four-photon ATI signal of fenchone (open squares), orientation of the C=O bond perpendicular to the light's propagation results in a negative PECD of --14.9\%, and for the angles $\beta \approx 55^\circ$ and $125^\circ$ it is large and positive (+14.9\%). Averaging over all orientations results in a small positive effect of +4.0\%. The three-photon PI signal of camphor illustrates the largest asymmetry of about +8\% for $\beta=90^\circ$ (solid circles in Fig.~\ref{fig:orient}). This asymmetry almost vanishes for small and large orientation angles, yielding somewhat smaller effect of +5.7\% on average. The four-photon ATI signal of camphor (open circles) exhibits a large positive effect of about +33\% for the perpendicular orientation $\beta=90^\circ$. Because the orientation of the C=O bond along the light propagation results in the negative effect of similar size (--33\% for $\beta=0^\circ$ and $180^\circ$), the average asymmetry drops to +7.4\%.

\begin{figure}
\includegraphics[scale=0.80]{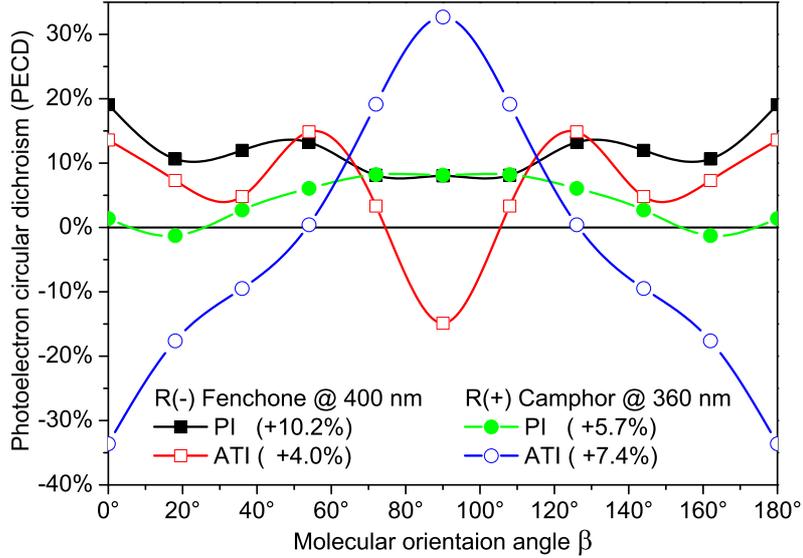}
\caption{Multiphoton PECDs (\ref{eq:mpiPECDbet}) as functions of the molecular orientation angle $\beta$ (between the C=O bond and the direction of the propagation of exciting pulses), computed for PI and ATI signals of fenchone and camphor exposed to laser pulses of different wavelength (see legends). An additional averaging of PECD over angle $\beta$ results in the total $\widetilde{PECD}$ listed in the legend in parenthesis (see also Tab.~\ref{tab:propt}).} \label{fig:orient}
\end{figure}

\section{Conclusions}
\label{sec:Summary}

The time-dependent single-center (TDSC) method, developed in paper~I, is applied to study multiphoton PECD of bicyclic ketones by short intense laser pulses. The present theory  reproduces the chiral asymmetry effects observed in Refs.~\cite{Lux12AngChm,Lux15CPC,LUXdiss,Lux16ATI,Kastner17wl} in the three-photon resonance-enhanced ionization (PI) and four-photon above-threshold ionization (ATI) spectra of randomly-oriented fenchone and camphor molecules. The overall sizes and signs of the chiral asymmetry effects, computed in the present work for the PI and ATI spectra of fenchone exposed to 400~nm laser pulses, are in a good agreement with the respective experimental results. Similarly to the experiments, we observe the change of sign of the three-photon PCED of fenchone  at shorter wavelengths. However, the effect computed in the present work for 380~nm laser pulses is somewhat stronger than the measured one. For camphor, we were able to reproduce the nodal structures, overall sizes and signs of PECD  effects observed in the experimental PI and ATI spectra at 400~nm. However, a good agrement between the theoretical and experimental results could be achieved  by using 360~nm laser pulses in the calculations.

The present work demonstrates the capability of the TDSC method to study multiphoton PECD effects in real chiral molecules. Although this method utilizes a single-active-electron approximation and a local $X\alpha$-potential for exchange interaction, it naturally incorporates  multiple scattering of the outgoing photoelectron wave on a chiral potential of the molecular ion, which is at the heart of PECD \cite{Ritchie3,Tia17,Hergenhahn04,Stener04}.
Our calculations provide additional evidences of the effect of an intermediate electronic state, which favors particular molecular orientations in the two-photon resonant excitation step \cite{Lehmann13jcp,Rafiee16wl,Kastner17wl,Goetz17}. It is demonstrated that a significant enhancement of the PECD, observed in Ref.~\cite{Tia17} in the one-photon ionization regime for a uniaxially-oriented chiral molecule, persists also in the multiphoton ionization regime. In the subsequent work, we plan to apply the TDSC method to theoretically study a passage from the multiphoton ionization of randomly-oriented fenchone molecules by intense UV/visible laser pulses to their strong-field tunnel ionization by infrared pulses in order to explore the universality of the PECD proposed in Ref.~\cite{Beaulieu16NJP}.

\begin{acknowledgements}
T. Baumert, C. Lux, A. Senftleben, H. Braun, A. Kastner and T. Ring are gratefully acknowledged for many valuable discussions. This work was supported by the Sonderforschungsbereich   SFB--1319 `Extreme light for sensing and driving molecular chirality' (subproject C1) of the Deutsche Forschungsgemeinschaft (DFG). Part of the calculations has been performed at the Lichtenberg-Hochleistungsrechner of the Technische Universit\"{a}t Darmstadt (Projects 0628 and 0660).
\end{acknowledgements}

\end{document}